\documentclass{aa}
\usepackage{epsfig,amsfonts,amssymb,graphicx,fancyheadings,caption,rotating}
\begin{document}
\def\lesim{\stackrel{<}{{}_{\sim}}} 

   \thesaurus{(08.19.1;  
                08.19.4;  
                04.19.1)} 
   \title{Type Ia supernova rate at $z \sim 0.1$
\thanks{This work is based on observations made at the
European Southern Observatory, La Silla, Chile.}
}

   \subtitle{}

   \author{D.~Hardin \inst{1,}$^*$,
C.~Afonso\inst{1}, 
C.~Alard\inst{9},
J.N.~Albert\inst{2},
A.~Amadon\inst{1},
J.~Andersen\inst{6},
R.~Ansari\inst{2}, 
\'E.~Aubourg\inst{1}, 
P.~Bareyre\inst{1,4}, 
F.~Bauer\inst{1},
J.P.~Beaulieu\inst{5},
G.~Blanc\inst{1},
A.~Bouquet\inst{4},
S.~Char${}^{\dag}$\inst{7},
X.~Charlot\inst{1},
F.~Couchot\inst{2}, 
C.~Coutures\inst{1}, 
F.~Derue\inst{2}, 
R.~Ferlet\inst{5},
J.F.~Glicenstein\inst{1},
B.~Goldman\inst{1},
A.~Gould\inst{1,8},
D.~Graff\inst{8},
M.~Gros\inst{1}, 
J.~Haissinski\inst{2}, 
J.C.~Hamilton\inst{4},
J.~de Kat\inst{1}, 
A.~Kim\inst{4},
T.~Lasserre\inst{1},
\'E.~Lesquoy\inst{1},
C.~Loup\inst{5},
C.~Magneville \inst{1}, 
B.~Mansoux\inst{2}, 
J.B.~Marquette\inst{5},
\'E.~Maurice\inst{3}, 
A.~Milsztajn \inst{1},  
M.~Moniez\inst{2},
N.~Palanque-Delabrouille\inst{1}, 
O.~Perdereau\inst{2},
L.~Pr\'evot\inst{3}, 
N.~Regnault\inst{2},
J.~Rich\inst{1}, 
M.~Spiro\inst{1},
A.~Vidal-Madjar\inst{5},
L.~Vigroux\inst{1},
S.~Zylberajch\inst{1}
\\   \indent   \indent
The EROS collaboration
}
\institute{
CEA, DSM, DAPNIA,
Centre d'\'Etudes de Saclay, 91191 Gif-sur-Yvette Cedex, France
\and
Laboratoire de l'Acc\'{e}l\'{e}rateur Lin\'{e}aire,
IN2P3 CNRS, Universit\'e de Paris-Sud, 91405 Orsay Cedex, France
\and
Observatoire de Marseille,
2 pl. Le Verrier, 13248 Marseille Cedex 04, France
\and
Coll\`ege de France, Physique Corpusculaire et Cosmologie, IN2P3-CNRS, 
11 pl. M. Berthelot, 75231 Paris Cedex, France
\and
Institut d'Astrophysique de Paris, INSU-CNRS,
98~bis Boulevard Arago, 75014 Paris, France
\and
Astronomical Observatory, Copenhagen University, Juliane Maries Vej 30, 
2100 Copenhagen, Denmark
\and
Universidad de la Serena, Facultad de Ciencias, Departamento de Fisica,
Casilla 554, La Serena, Chile
\and
Departments of Astronomy and Physics, Ohio State University, Columbus, 
OH 43210, U.S.A.
\and
DASGAL, 77 avenue de l'Observatoire, 75014 Paris, France\\
\hspace*{-4.04mm} $^*\,$ Now at LPNHE, Universit\'e Paris VI, 4 place
Jussieu, F-75252 Paris Cedex 05, France }
\offprints{D. Hardin (hardin@in2p3.fr)}
  \date{Received ; accepted}
 \authorrunning{D. Hardin et al.}
  \maketitle
\def\lsun {{ \rm \, L_\odot}} 
\def\lbsun {{\lsun}_B} 
\def\lvsun {{\lsun}_V} 

   \begin{abstract}
     We present the EROS nearby supernova
($z \sim 0.02 - 0.2$) search  and the analysis of the
first year of data (1997). 
A total of $80$ square degrees were surveyed. 
Eight supernov{\ae} were detected, 
four of which were spectroscopically identified as type Ia
supernov{\ae}. 
The search efficiency was determined with
a Monte-Carlo simulation  taking into account 
the efficiencies for both supernova detection  
and  host galaxy identification. 
Assuming that for a given galaxy  the supernova rate 
is proportional to the galactic luminosity, we compute
a type 
Ia supernova explosion rate of:
${\cal R } =  0.44  {}_{-0.21}^{+0.35} \, {}_{-0.07}^{+0.13} \,   h^2 
  \: / 10^{10} \, \lbsun /  \,100 \,{\rm yrs}$  
at an average redshift of $\sim 0.1$ where the errors are
respectively statistical and systematic  (type misidentification included).
\end{abstract}

%

\section{Introduction}

Type Ia supernov{\ae} have been shown to be powerful cosmological
distance indicators useful for the determination of
the expansion rate $H_{0}$ and the density parameters
for matter and vacuum energy $\Omega_{M}$ and  $\Omega_{\Lambda}$ 
(Riess et al. \cite{RIE98}; Perlmutter et al. \cite{PER99}).
As such, they are the subject of intense study and
a  number of nearby and distant supernova searches have been launched.
Studies of type Ia light-curve shapes, peak magnitudes,
colors and spectra have been performed in order to get better 
insight into the nature  of these exploding stars.

Supernova rates, and in particular their evolution as a function of
redshift, is another probe provided by  supernova science.
Not only are these rates important for tracking the chemical evolution
of the universe, but as a tracer of stellar evolution,  they also 
contain information on which type of progenitor system
produces type Ia supernov{\ae}. 
If the evolution of the supernova rate is understood, they
may also be used to determine the cosmological parameters
 $\Omega_{M}$ and $\Omega_{\Lambda}$ via the number count-redshift
relationship.
With enough statistics, it is possible to measure the supernova luminosity
function which is of critical importance for understanding Malmquist bias.

The supernova rate has been measured at low redshift in searches
based on  visual scanning
of  photographic plates (Cappellaro  et al.  \cite{CAP97}). 
At high redshift the rate has been measured  with automatic
subtraction of CCD  images  (Pain  et al.  \cite{PAI96}).
The supernova rate for nearby galaxies has previously been
estimated using CCD data by Muller  et al.  (\cite{MUL92}).  
Here we report
the first measurement of the rate for low redshift galaxies
using CCDs and with a full numerical
calculation of the detection efficiency. 
The search was performed by EROS  
(Exp\'erience de Recherche d'Objets Sombres).

In 1990 EROS 
  engaged in a search for massive compact halo objects
(MACHOs) via gravitational microlensing. 
Since 1996 EROS has been using  a dedicated 1 meter telescope, the  Marly telescope, 
   at the ESO La Silla
observatory.
The telescope is equipped with  
two  detector mosaics, each consisting  of eight  2048$\times$2048 pixel CCDs
covering  a $0.7 \times 1.4$ ${\rm deg}^{2}$ field.
A dichroic cube allows the simultaneous recording of two images
in  wide  ($\Delta \lambda \sim 200$ nm) 
non-standard  red ($\bar{\lambda} \sim 760$ nm) and  visible 
($\bar{\lambda} \sim 560$ nm) bands
hereafter referred to as  $R_{\rm Eros}$ and $V_{\rm Eros}$. The telescope, 
camera and  telescope operation
are described in Bauer et al. (\cite{BAU97}).

The greater part of EROS observing time  is devoted to the search 
for microlensing events by observing the Magellanic 
Clouds, the Galactic Bulge and the Galactic Disk 
(see {\em e.g.} Palanque-Delabrouille  et al.  
 \cite{PAL98}; Derue  et al.  \cite{DER99}). 
However, EROS also devotes 10\% of its observing time 
to the observation of fields at high Galactic latitudes 
in order to search for supernov{\ae} at low redshifts and 
to measure stellar proper motions. 

The EROS supernova search aims at discovering a homogeneous 
sample of supernov{\ae} in the low redshift range.
Our main scientific goals are the derivation of the nearby 
supernova rates, and the study of the empirical correlation of
the peak luminosity with the light-curve shape.
EROS has discovered more than 60 supernov{\ae} since 1997.

In this paper, we present a measurement of the  type Ia supernova 
explosion rate.  Because of backgrounds due to variable
stars and asteroids, we restricted the search to identified
galaxies.  To derive a total supernova rate, we 
therefore assume
that the rate for a given galaxy is proportional to the
galactic luminosity.  For this low statistic supernova sample, no
attempt is made to derive rates for different galaxy types.

In Section 2, we give the details of our supernova
search that are relevant for the rate determination.
In Section 3  we present the
subset of data considered in this analysis and  the
supernov{\ae} we found.
The galaxy sample for the supernova search is described in
Section 4.
In Section 5
we describe the determination of our detection efficiency using a Monte Carlo
simulation and derive the supernova rate.
Finally, in section 6 we present a discussion and interpretation of
our results. More details on this measurement can be found in Hardin
(\cite{HAR98}).

\section{The search technique}
\label{sec1}

\begin{figure}
\begin{center}
\psfig{file=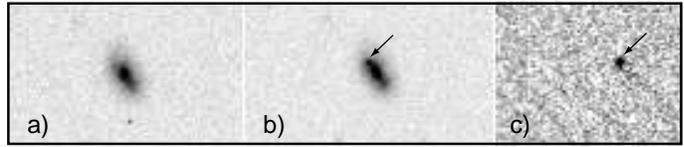,width=.5\textwidth}
\caption{Illustration of the detection technique (detection images of 
SN 1997eb). a) Reference image taken on October 7, 1997. 
b) Detection image taken on November 7, 1997. 
The supernova can barely be distinguished
 from the galaxy nucleus.  c)  The subtracted frame, obtained after
geometric and  photometric alignment of a) and b) frames. 
The stellar PSFs  are matched by convolving 
the best image. The supernova now stands alone.}
\label{rcs}
\end{center}
\end{figure}

We searched for supernov{\ae} by comparing an image 
of a given field with a reference image of the same 
field taken at least two weeks before. The images are taken during 
dark time.
Only the visible band images are used for the search, but we use 
both bands when computing the absolute calibration of the images.
The technique is illustrated in Fig. \ref{rcs}.
Each observation consists 
of two five minute exposures which are combined 
to form a ten minute exposure after 
identification of cosmic rays.
Five to ten fields were observed each night.
For the observations used in this paper, calibration
fields  (Landolt \cite{LAN92})  were also
observed before and after the series of supernova fields.
This allowed us to calibrate our non-standard filters
and to eliminate non-photometric nights from the
rate measurement.

The automated processing permits the remote analysis in France
on computers set up at the La Silla observatory of up 
to 20 deg$^{2}$ during the day following the night of  observation. 
Since the processing is monitored
from France, it requires robustness and a limited volume of
output.

 The method for comparing the reference and the search
images is as follows.
The reference image is 
aligned geometrically and photometrically with the 
search image, 
and the image of superior seeing is convolved to match 
the point spread function (PSF) of the other image. 
The two images are then subtracted and
the frame thus obtained is searched for star-like sources.
The selection of the candidate supernov{\ae} 
is performed by applying a set of cuts 
on the objects   detected  on the subtracted frame.
These cuts are tuned with a Monte-Carlo simulation.

The first and most important cut is applied on the total flux of the candidate,
and is equivalent to a limiting magnitude 
of $V_J \sim 21.5$.
Afterwards, a series of cuts involving the comparison of 
 the candidate shape 
({\em i.e.} the second moments of a Gaussian fit) with 
that of the frame stars
is designed to eliminate inaccurate subtractions. 
Finally, the detection threshold is set slightly higher within 
1.8 arcsec of   the 
center of bright galaxies.

The treatment ends with the visual scanning of 
$\sim 5 $ candidates per square degree by two independent
observers.
Most of the candidates are 
residual cosmic rays, subtraction artifacts, asteroids, and variable stars.
The residual cosmic rays  are identified by the fact that they only appear on 
one of the two   five minute
exposures. Asteroids and variable stars are eliminated by 
requiring the candidate to appear in  or near a host identified as a galaxy. 
As a consequence of this last requirement, 
no supernov{\ae} with an undetected host can be found. This is taken 
into account in the calculation of the search efficiency.

\section{The data set}
\label{sec2}
\begin{table*}
\begin{center}
\begin{tabular}{|c|c|c|c||c|c|}\hline
zone & \# fields & $\alpha$ (J2000)   & $\delta$  (J2000)  & $l$ & $b$ \\ \hline
\multicolumn{6}{|c|}{October 1997 run}\\ \hline
Southern hem.  & 43  & 
 $22 h 30'  \lesim \alpha  \lesim \:2 h 30' $ & $ \delta \sim -40^{\circ}$ & $250^{\circ}  \lesim l  \lesim 360^{\circ} $ & $ -75^{\circ} \lesim b \lesim -60^{\circ}$ \\\hline
Abell Cluster  & 8  & $2 h 00'  \lesim \alpha  \lesim \:3 h 00' $ & $ \delta \sim 0^{\circ}$ 
 & $150^{\circ}  \lesim l  \lesim 180^{\circ} $ & $ -50^{\circ} \lesim b \lesim -60^{\circ}$ \\\hline
\multicolumn{6}{|c|}{November 1997 run}\\ \hline
Abell Cluster  & 4 & $2 h 00'  \lesim \alpha  \lesim \:3 h 00' $ & $ \delta \sim 0^{\circ}$ 
 & $150^{\circ}  \lesim l  \lesim 180^{\circ} $ & $ -50^{\circ} \lesim b \lesim -60^{\circ}$ \\\hline
Northern hem.  & 36 & 
 $3 h 00'  \lesim \alpha  \lesim \:4 h 00' $ & $ \delta \sim 0^{\circ}$ 
 & $175^{\circ}  \lesim l  \lesim 190^{\circ} $ & $ -35^{\circ} \lesim b \lesim -55^{\circ}$ \\\hline\hline
\end{tabular}
\end{center}
\caption{List of the surveyed sky zones during the October  and the November 
1997 runs. The number of fields surveyed in each zone is given in column 2. 
Due to a temporary malfunction of CCD \# 7 on the camera corresponding
to the visible band, one field
only covers  $0.871$  deg$^2$. The total surveyed area is thus 
$ \sim 45$ (resp. $ \sim 35$) square degrees 
for the October (resp. November) 1997 run.
The southern hemisphere  fields 
are included in the Las Campanas Redshift Survey fields.}
\label{tabchp}
\end{table*}
\begin{table*}
\begin{center}
\begin{tabular}{||lc|c|c|c|c|c|c||} \hline 
\multicolumn{2}{||c|}{IAU name} &  date   &    type  &    $z$  & host  & discovery   &   dist. from \\ 
& & & & & $R_c$ mag. & $V_J$ mag & host core  \\ \hline \hline 
SN 1997dh &(1)& {\small 20/10/1997}  & Ic  &  0.05  
& $15.7$ & $20.$ & 5.0 '' \\ \hline
SN 1997dk &(2)& 26/10/1997  &  Ia  &  0.05  & $15.2$ & $20.$
& $11.0$ ''  \\ \hline
SN 1997dl &(2)& 26/10/1997  &  Ia  &   0.05 & $15.2$  & $19.5$ 
& $11.0$ ''  \\ \hline
SN 1997dm &(2)& 26/10/1997  &  IIp (l.c.) &  0.03 & $14.4$  
& $21.$  & $6.7$ ''\\ \hline
SN 1997eb& (3)& 19/11/1997    &  II (l.c.)  &   0.08 &  $17.1 $ &  $21. $5 
 & $2.9$ '' \\ \hline
 SN 1997ec &(3)& 20/11/1997  &   IIp (l.c.)  &  0.12 &  $16.3 $ 
&   $21.5 $   & $3.7$ ''  \\ \hline
SN 1997ed& (3)& 22/11/1997  &  Ia  & 0.15  & $ 17.5$ & $21.$
& $2.8$ ''    \\ \hline
SN 1997ee& (3) & 23/11/1997     &  Ia  &   0.17  & $17.0$ &  $21.$
& $8.6$ ''  \\ \hline
\end{tabular}
\end{center}
\caption{Main characteristics of the  supernov{\ae} detected during
the October  and the November 1997 runs. In the third column, (l.c.) 
indicates that the supernova type was derived from  
its light curve. 
All magnitudes are corrected for Galactic absorption. Note that SN 1997dk and SN 1997dl occurred
in the same galaxy, 2 months apart according to their spectra.
The references are as follows:
(1) IAU6760, \cite{IAU6760}; (2) IAU6762, \cite{IAU6762}; (3) IAU6782,  \cite{IAU6782}, and IAU6785,  \cite{IAU6785}. }
\label{tabsn}
\end{table*}
For this analysis, we have used the data obtained during the two 
search runs conducted during October and November 1997.
A description of the observed  fields is given in Tab. \ref{tabchp}.
They have been  divided  into 3 zones. The southern hemisphere
fields have been chosen far from the Galactic plane
(mean Galactic latitude  $<\!b\!> \sim -70^{\circ}$), in a region 
covered by The Las Campanas Redshift Survey (LCRS, 
Shectman  et al.  \cite{SHE96}). 
 While the Abell Cluster fields\footnote{Our original program 
included a study of the peculiar velocities of these clusters
(Reiss et al., \cite{REI98}).
 are centered on Abell Clusters at a mean redshift
of $z \sim 0.17$, the search in these fields was not restricted to the
galaxies that belong to these clusters. }
The northern hemisphere 
fields have 
been chosen so as to be observable from
the La Silla (Chile) and Apache Point (New Mexico, USA) observatories
at the time of the November 1997 run. As a consequence, their mean 
Galactic latitude ($<\!b\!> \sim -50^{\circ}$) is not as high  as
in the southern hemisphere zone. The total area of sky 
(80  square degrees) was covered over a period of  10 days.

Among the 8 detected supernov{\ae},  5 could be classified
according to their spectra.
The type of the three remaining supernov{\ae} was obtained using the
shapes of their light curves. 
The characteristics of these eight 
supernov{\ae} are summarized in Tab. \ref{tabsn}.


\section{The galaxy sample}
\label{sec3}

In order to determine the supernova rate, we must have
a well-defined sample of potential host galaxies 
with a known distribution of redshifts.
In this section, we describe our galaxy identification
and redshift determination algorithms and their verification
using LCRS galaxies in our fields.

\subsection{Galaxy identification} 
The position and flux of the sources on the reference 
and detection frames
are measured with the software
for source extraction SExtractor (Bertin \& Arnouts  \cite{BER96}). 
SExtractor measures the total flux of extended sources,
and provides in addition a star-galaxy separation estimator, 
{\sc class star}, which uses the 
intensity profile. 

Our
 galaxy selection criterion is based on the (total) 
$V_{\rm Eros}$ magnitude
and {\sc class star}. Because of the PSF variation
over the CCD mosaic, the cut applied on {\sc class star} differs 
from one CCD to the other. On the other hand, the cut on 
$V_{\rm Eros}$ is set equal for all frames and corresponds to a 
magnitude ${R_c}_0 \sim 18.7$ for the October 1997  search fields, 
and ${R_c}_0 \sim 18.4$
 for  the November 1997 search fields. This difference is due 
to  Galactic absorption. The cut on the galaxy magnitude was to ensure 
that the host was classified
as such during the visual scanning. 
 At $z=0.1$, this corresponds to an absolute magnitude of 
$M_{R_c} \sim -20. + 5 
\log(H_0/60 \, {\rm km}{\rm s}^{-1}{\rm Mpc}^{-1})$ for the
host galaxy.

\begin{figure}
\begin{center}
\psfig{file=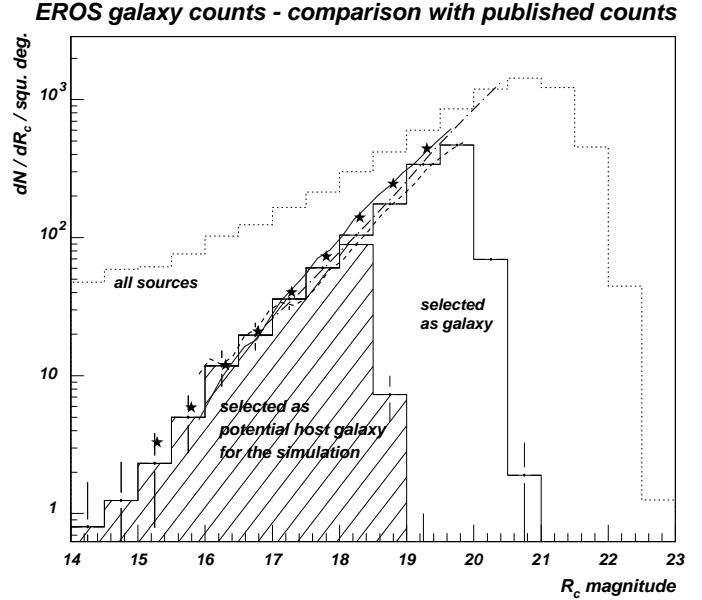,width=.5\textwidth}
\caption{Comparison of EROS galaxy counts (histograms)
 with counts from Bertin \& Dennefeld  (\cite{BER97}; stars), Weir  et al.  
(\cite{WEI95}; 
bold dash-dot line), and the northern (bold full line) and southern counts
(bold dashed line)  from Picard (\cite{PIC91}). 
The dotted line histogram shows all the detected 
sources on the frames, both stars and galaxies, the full line histogram  
 the objects selected as galaxies (with $R_c \lesim 20.2$), 
and the hatched  histogram 
the objects selected as potential hosts for the simulation, 
{\em i.e.} with a magnitude $R_c \lesim 18.7$. 
This corresponds to a solid angle of 
38 deg$^2$ (October 1997 data). 
There is good 
agreement between our galaxy counts and published counts.}
\label{galcount}
\end{center}
\end{figure}

This selection criterion was first tested on the fields 
shared with the LCRS: 90\% of the LCRS redshift catalog galaxies
were identified as such by our criterion, and 95\% of the selected 
objects with a magnitude $R_c \lesim 17.5$ (LCRS completeness limit)
 were LCRS galaxies.

Having identified the galaxies on our frames and \mbox{derived} their $R_c$ 
(total) magnitude from their $V_{\rm Eros}$ and
 $R_{\rm Eros}$  magnitudes and the  calibration (see section 2), 
we obtain a galaxy count per magnitude
and per deg$^2$, which we can compare with  published counts. 
The agreement is illustrated in Fig. \ref{galcount}. 
One can also identify the complementary set of objects
as stars and obtain in this way star counts as a function
of magnitude.  These star counts compare well with the Galactic
star-count model of 
Bahcall \& Soneira  (\cite{BAH84}).

At this stage, $\sim 225$ objects per  deg$^2$ 
are selected as potential  host galaxies.

\subsection{Magnitude-redshift relationship} 

For a galaxy of a given magnitude
 $R_{\rm c \: gal}$, the  probability that its 
redshift is $z$ is given by :
\begin{eqnarray} \label{problaw}
p ( z |R_{\rm c \: gal})    &  
\propto   & \frac{dV_c}{\!\!\!\!dz}( z ) \times \\ 
\nonumber
& &\frac{dN_{\rm gal}}{\!\!\!\!dM}\left(M 
= R_{\rm c \: gal} - \mu_L( z ) 
- K_{\rm gal}( z ) \right) 
\end{eqnarray}
In this equation, 
$dV_c/dz$ is the derivative of the comoving volume with respect to  $z$, and
$\mu_L$ is the luminosity distance modulus.
The cosmological  parameters $(\Omega_{M};\Omega_{\Lambda})$ are set to
the value $(0.3;0.)$. Taking the  values
$(\Omega_{M};\Omega_{\Lambda})=(0.3;0.7)$
based on  distant type Ia supernov{\ae}
(Riess et al. \cite{RIE98}; Perlmutter et al. \cite{PER99}) 
decreases  the computed rate by less than 5\%. 

The distribution of the number of galaxies per 
unit absolute magnitude $dN_{\rm gal}/{dM}$  is the Schechter law:
\begin{equation}\label{sche}
\frac{dN_{\rm gal}}{\!\!\!\!dM} \propto  
 10^{-0.4 \times (\alpha + 1 ) \times (M - M_{\ast})} \:
\exp\left(- 10^{-0.4 \times (M - M_{\ast})} \right)
\end{equation}
measured  by the LCRS
(Lin  et al.  \cite{LIN96}) where $\alpha_{\rm LCRS}  =  -0.7 $ and
${M_{\ast}}_{ \: R_c \: {\rm LCRS}}  =  -20.29 + 5 \log h$
($h=H_0/100 \, {\rm km}{\rm s}^{-1}{\rm Mpc}^{-1}$).

The K-correction is simply given by $K_{\rm gal}( z ) = 2.5 \log(1+z)$
as  proposed by Lin  et al.  (\cite{LIN96}).

For a galaxy magnitude of $R_{\rm c \: gal} \simeq 17$, 
 the mean and rms value of the $z$-distribution thus obtained 
are respectively  $\sim 0.1$ and $\sim 0.05$.

The model was checked  by comparing,
for each galaxy with an LCRS redshift, the mean of $p(z|R_{\rm c \: gal})$ with
the measured LCRS redshift.
Satisfactory agreement was obtained after adding 0.25 to  
the EROS total magnitudes (as measured with SExtractor) 
so as to align them with the isophotal
LCRS magnitudes.


\section{The detection efficiency and the supernova rate}
 \label{sec4}

The supernova rate ${\cal R}$ in the
rest frame  is expressed in SNu, {\em i.e.} in supernov{\ae} per unit
time per unit blue luminosity, $1 \,  {\rm SNu} = 1 \, {\rm SN} 
\: / 10^{10} \, \lbsun /  \,100 \,{\rm yrs}$.
The number of supernov{\ae} ${\cal N}$ 
expected to be detected 
is given by  a sum over
the observed galaxies $i$ weighted by their blue luminosities $L_{i}$:
\begin{eqnarray}
{\cal N} & = & {\cal R} \times {\cal S}, \label{NRS} \\
{\cal S} & = & \sum_{\mbox{\tiny gal. i}}
L_{i}  \,\, \int_{-\infty}^{\infty} \, \epsilon_{i} (t,z_{i})\; dt
\nonumber
\end{eqnarray}
where $\epsilon_{i} (t,z_{i})$ is  the efficiency to detect
in galaxy $i$ 
a type Ia supernova whose maximum occurs  at time $t$  
 in the supernova rest frame. 
The efficiency clearly depends on the galaxy redshift $z_{i}$ but
also on other factors  such as the galactic luminosity and surface-brightness
profile.
The integral $\int\epsilon_{i} dt$ is sometimes called the ``control time''
for the galaxy $i$.

We now need to determine the sum ${\cal S}$ so the
rate can be evaluated through Eq. \ref{NRS}.
We do not know the luminosities of most of  the observed
galaxies because most of their redshifts are not known.
As discussed in the previous section, 
we do, however, know the probability $p ( z |R_{\rm c \: gal})$ that the
galaxy of magnitude $R_{\rm c \: gal}$ 
has a redshift $z$.  
We therefore replace the above expression for ${\cal S}$ with
\begin{equation}
{\cal S}  =  \sum_{\mbox{\tiny gal. i}} \, \int_{0}^{\infty} dz \; 
p(z|R_{c\:i}) \, L_{i}(R_{c\:i},z)\,
\int_{-\infty}^{\infty}  \epsilon_{i} (t,z)\; dt
\end{equation}
where the blue luminosity $L_i$ is computed using the magnitude $R_{c\:i}$
and the redshift $z$.
The double integral in this equation was evaluated for each galaxy
by Monte Carlo simulation of supernov{\ae} that are subjected to the
same detection procedures as the real data.

Because there is a better match between the 
  EROS visible band  
and the standard V filter than with  the standard B filter, 
the absolute luminosities of galaxies 
are first computed in the standard  V band, and then converted into  
 absolute luminosities in the standard B band. 
We use the difference between the solar
$<\!\!B\!\!-\!\!V\!\!>$ colour: $ <\!\!B\!\!-\!\!V\!\!>_{\odot} = 0.65$ and a mean $<\!\!B\!\!-\!\!V\!\!>$ colour
for galaxies, computed from the results of Fioc \& Rocca-Volmerange  (\cite{FIO97}): 
$ <\!\!B\!\!-\!\!V\!\!>_{\rm gal} = 0.77$.

\subsection{Supernova simulation}

The Monte Carlo simulation proceeds as follows.
For each potential host galaxy, a series of redshifts 
is drawn randomly according
to the distribution appropriate for the galaxy apparent magnitude.
For each redshift, 
a  time $t$ since maximum 
is drawn uniformly between $-15$ and $+120$ days
 in the supernova rest frame. The 
absolute magnitudes $M_{V_J}$ and $M_{R_c}$ 
of the supernova at that time are taken from the
templates in these standard filters provided by 
Riess  et al.  (\cite{RIE96}).
To account for the observed spread of supernova
luminosities, a gaussian scatter
$\Delta M = 0.2 \, {\rm mag}$ was added.  
The apparent magnitude  of the simulated supernova is computed 
from the redshift  
using the  K-correction from Nugent \& Kim (\cite{NUG99})
and assuming $(\Omega_M, \Omega_{\Lambda})=(0.3,0.)$ as above. 

The supernova standard apparent magnitudes are then converted into a
$V_{\rm Eros}$ flux using our calibration.
The position of the supernova in the 
host galaxy is drawn according to a two-dimensional 
Gaussian distribution, 
the first and  second order moments and orientation of which
are those of the host object.
 
As supernov{\ae} are stellar sources,  they appear on the frame 
with the same PSF as the field stars do. The PSF is modeled by a 
two-dimension integrated Gaussian. \mbox{Its} second order moments are 
set to the mean moments computed by fitting the stars on 
the same  CCD  quadrant, in
order to take into account the spatial variation of the PSF.

\subsection{Detection efficiency}

The detection efficiency is derived by comparing the list of 
simulated supernov{\ae} and the list of the 
supernov{\ae} selected by the search cuts.
\begin{figure}
\begin{center}
\psfig{file=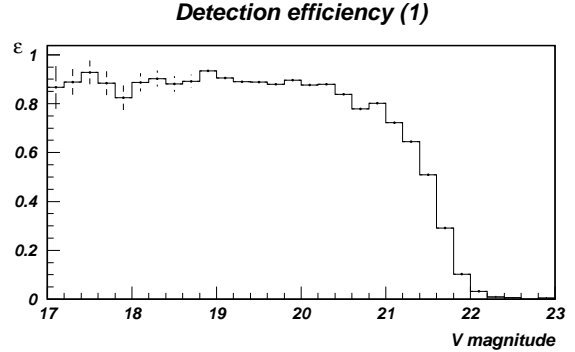,width=.5\textwidth}
\vspace{-1cm}
\caption{The detection efficiency $\varepsilon$ as a function of the simulated 
supernova magnitude $V_J$ at discovery. 
It corresponds to a limiting magnitude
$V_J \sim  21.5$. 
The maximum efficiency is below $1$ because of the
masked areas near CCD defects and saturated stars. }
\label{effic_mag}
\end{center}
\end{figure}

\begin{figure}
\begin{center}
\psfig{file=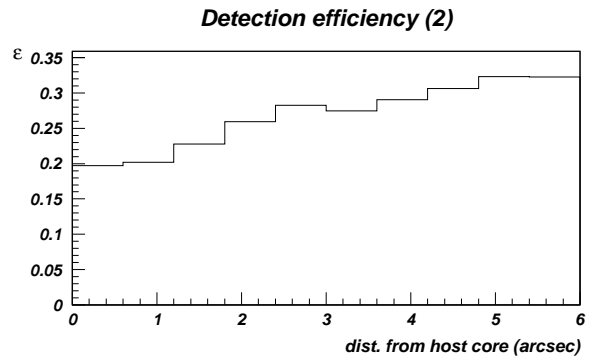,width=.5\textwidth}
\vspace{-1cm}
\caption{Dependence of the 
detection efficiency  $\varepsilon$ on the distance between the
simulated supernova and its host galaxy core.
The average efficiency is as low as $\sim$30\% because supernova light curves
are sampled  up to $120$ days after maximum, causing 
$\sim 75$\% of them to drop below
$V_J =22$.}
\label{effic_dist}
\end{center}
\end{figure}

\begin{figure}
\begin{center}
\psfig{file=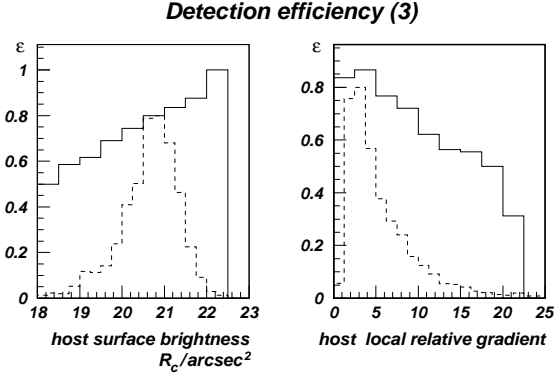,width=.5\textwidth}
\caption{Detection efficiency  $\varepsilon$ as a function of  the host 
galaxy surface brightness (left pannel)
and  of the local gradient at the point where the 
simulated supernova is added (right pannel).
The local gradient was taken to be
$ g  = {\rm max}\left({\rm pixel}(x_i,y_i) -{\rm pixel}(x_j,y_j) \right)_{|j-i|\leq 1}/B$
where B is the local Poisson noise.
The distributions of the surface brightness and the local gradient 
are superimposed (dashed lines).
The distributions are shown for
supernov{\ae} whose  magnitude is in the range
$20.5<V_J<21$.}
\label{effic_div}
\end{center}
\end{figure}

The detection efficiency as a function of magnitude is shown 
in Fig. \ref{effic_mag}. It appears as a smoothed step function, 
corresponding to a limiting magnitude $V_J \sim 21.5$.
There is a slight dependence on the distance
of the supernova from the host galaxy core, as shown in 
Fig. \ref{effic_dist}.

As foreseen, the detection efficiency depends on the host 
galaxy characteristics, such as the surface brightness
 or the local gradient where the 
simulated supernova is added. 
These results are
presented in Fig. \ref{effic_div}, where
the distribution of these two parameters have been superimposed 
on the efficiency. The distributions are shown for
supernov{\ae} whose magnitude is in the range
$20.5<V_J<21$. They peak where the efficiency is good,  
thus moderating the impact of the efficiency behaviour.

\subsection{Redshift distribution}

\begin{figure}
\begin{center}
\psfig{file=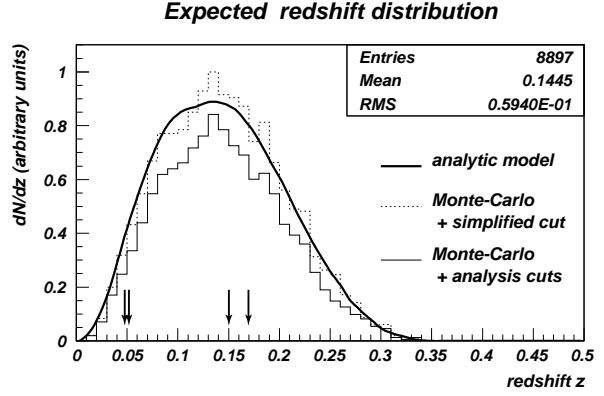,width=.5\textwidth}
\vspace{-2cm}
\caption{The expected redshift distribution. The thin solid line
shows the distribution
 in $z$  derived
from the Monte-Carlo simulation when the analysis cuts are applied. 
The dotted line shows the same but with
 only two cuts   applied, one  on the supernova 
magnitude ($V_J < 21.5$) and  the other one on the host galaxy $V_{\rm Eros}$
magnitude equivalent to $R_{c \, {\rm gal}}\lesim 18.5$. 
The latter distribution compares well with a simple analytic model 
as described in the text (thick solid line).
The arrows indicate the redshift values 
of the 4 discovered  type Ia supernov{\ae}.}
\label{dist_z}
\end{center}
\end{figure}

The Monte-Carlo computed 
redshift distribution of detected supernov{\ae} 
is shown
in Fig. \ref{dist_z}. 
The thin solid line shows  the distribution when all the
analysis cuts are applied. 
For comparison,   the dotted line shows the redshift distribution when only two cuts
are applied, one on the supernova 
magnitude $V_J < V_{J_0} = 21.5$, and one on the host galaxy $V_{\rm Eros}$
magnitude equivalent to $R_{c \, {\rm gal}} <  {R_{c \, {\rm gal}}}_0 =18.7$.
 The thick solid line shows the distribution obtained with an analytic
model where we assume that the type Ia supernova rate is proportionnal
to the host galaxy luminosity and that the galaxy luminosity distribution
is given by the LCRS Schechter law (Eq. 2). In this analytic model the cuts
are described by two step-functions whose thresholds are respectively
$V_{J_0}$ and ${R_{c \, {\rm gal}}}_0$.
Using this analytic model, we find that
imposing the host galaxy magnitude cut reduces
the number of detected supernova by a factor $\sim 2.3$.

The  observed redshift distribution of the 4 
discovered type Ia supernov{\ae}
is in agreement  with the Monte-Carlo distribution (the Kolmogorov-Smirnov 
compatibility test gives a 25\% probability), 
although the small-number statistics prevents any definitive conclusion.
The mean and rms of the expected redshifts, respectively $0.14$ and $0.06$,
are  to be compared with  the mean and rms of the observed  redshifts, 
respectively $0.10$ and $0.06$.

\subsection{Explosion rate}

The rate is computed by dividing
the  number of observed supernov{\ae} ${\cal N}_{\rm obs}$ 
by the sum ${\cal S}$:
\begin{eqnarray}
{\cal R} & = & {\cal N}_{\rm obs} / {\cal S}
\end{eqnarray}

The sum obtained is ${\cal S}= 9.09 \, 10^{12} \, h^{-2} \, \lbsun \, {\rm yr}$.  
For ${\cal N}_{\rm obs} = 4 $ type Ia supernov{\ae},  
this corresponds to a rate
${\cal R} = 0.44 {}_{-0.21}^{+0.35} 
\, h^{2} \, / 10^{10} \, \lbsun / \,100  \,{\rm yr}$ 
where the error is statistical and given at a 68\% confidence level.


\section{Discussion}

The numerical calculation of the efficiency makes it 
possible to study various systematic effects. We study 
here the  effects inherent in the hypotheses
used in the Monte Carlo simulation, such as
the calibration relations  or 
the assumed type Ia supernova light-curve shape.

\subsection{Calibration}

The calibration enters the calculation at three different steps:
when measuring the galaxy magnitude $R_c$, which is used to 
draw its redshift, when computing the ADU flux 
of a simulated supernova as a function of
its $V_J$ and  $R_c$ magnitudes, and when estimating 
the galaxy absolute luminosity in the $V_J$ filter
 for the rate computation in SNu. 
The redshift galaxy 
distribution is  verified with the LCRS galaxies,
therefore we shall
only take into account here the last two steps.
Their contributions are of opposite sign, so that the 
estimated 10\% uncertainty  in the calibration zero-point yields
only a 5\% uncertainty in the computed rate.

\subsection{Galaxy counts and redshift distribution}

To take into account the requirement for the detected supernova 
to be found in a galaxy, potential host galaxies were selected 
by requiring that their magnitude $R_c$ satisfy  $R_c < {R_c}_0$
(section 4.1). A 0.2 magnitude shift on ${R_c}_0$
translates into a 10\% shift on the rate. 

The galaxy sample misidentification is estimated  to  
be about 10\%
(section 4.1), which gives 
rise to a 10\% uncertainty in the computed rate.

\subsection{Supernova luminosity distribution}

In the simulation, the supernova light curve was approximated 
by  $V_J$ and $R_c$ templates of Riess  et al.  (\cite{RIE96}). 
The light-curve shape is described 
through two parameters: the peak magnitude $M_{V_J \,{\rm max}}$ and 
the stretch factor $s$ (Perlmutter  et al.  \cite{PER99}).
We assumed in the analysis that $M_{V_J \,{\rm max}} = -18.4 + 5 \log h$ with an added gaussian 
scatter of $0.2$ mag, which corresponds to the observed distribution
for the 16 supernov{\ae} from  the Hamuy  et al.  (\cite{HAM96}) 
sample reanalysed 
in Perlmutter  et al.  (\cite{PER99}). No stretch factor correction 
was taken into account,  {\em i.e.} $s$ was set to $s=1$.
To test the robustness of this analysis, we reperformed the analysis
assuming, following Perlmutter  et al.  (\cite{PER99}),
that the peak magnitude is described by 
$M_{V_J \,{\rm max}} = M_{{V_J}_0 \,{\rm max}} -0.6\times(s-1)$, 
where the $M_{{V_J}_0 \,{\rm max}}$ distribution is a Gaussian 
centered on $M_{V_J} = -18.4 + 5 \log h$ with a scatter
$\sigma = 0.2$ mag., and $s$ is a Gaussian centered on $s=1$
with  a scatter
$\sigma = 0.1$. 
The rate calculated with these assumptions differed from that
calculated with the standard assumptions by less than 1\%.

On the other hand, it seems possible that the type Ia supernova
 mean  peak
magnitude might be uncertain to about 0.1 mag.
A shift of $0.1$ mag in the peak magnitude results in a 
5\% variation of the rate.

\subsection{Total number of detected type Ia}
The type  of SN 1997eb is unclear.
This is taken into account by adding $ {}_{-0.00}^{+0.11} \, {\rm SNu}$ 
to the systematic uncertainty.

\section{Conclusion}

\begin{table}
\begin{center}
 \caption{Summary of systematic errors.}
\label{err}
\begin{tabular}{|l|r|}
\hline\hline
Magnitude calibration & $\pm\,$ 5\%  \\\hline
Limiting magnitude ${R_c}_0$ for the & \\
galaxy selection & $\pm$10\%  \\\hline
Galaxy misidentification & $\pm$10\%  \\\hline
SN mean peak magnitude  & $\pm\,$ 5\%  \\\hline
&\\
SN type  misidentification & ${}^{\mbox{ +25}}_{\mbox{\,\,\,-- \hspace{0.05cm}$\!$ 0}}$\%\\
&\\\hline
&\\
Total systematic error & ${}^{\mbox{ +30}}_{\mbox{\,\,\,--\,16}}$\%\\
&\\
\hline\hline
\end{tabular}
\end{center}
\end{table}

We measure a rate of:
$${\cal R } =  0.44  {}_{-0.21}^{+0.35} \, {}_{-0.07}^{+0.13} \,   h^2 
\, \, \, {\rm SNu}$$
where  the systematic errors from Tab. \ref{err} have been added in
quadrature.
The statistical error 
corresponds to a 68\% confidence level.
This rate compares well with the rate
derived by Cappellaro  et al.  (\cite{CAP99})
from the combined sample of five nearby
photographic searches,  ${\cal R } =  0.35 \pm 0.11 \, h^2$ SNu
at a mean redshift of $z \sim 0.01$.
Similarly, our rate compares well with the Supernova Cosmology Project 
rate for type Ia at $z \sim 0.4$ derived by Pain  et al.  (\cite{PAI96}),
${\cal R } =  0.82^{+0.65}_{-0.45}\, h^2$ SNu.

The rate can also be expressed in $h^3 / {\rm Mpc}^{3} / {\rm yr}$
by multiplying the rate in SNu by the luminous density of
the universe $\rho_L  = 1.4 \pm 0.1 \, 10^{8}\, h \, \lsun \,
 {\rm Mpc}^{-3}$ 
(Lin  et al.  \cite{LIN96}).
This gives:
$${\cal R } = 0.62 \pm {}_{-0.29}^{+0.49} {}_{-0.11}^{+0.19} \, 10^{-4} \, h^3 / {\rm Mpc}^{3} / {\rm yr.}$$

\begin{acknowledgements}
We are grateful to D. Lacroix and the technical staff at the Observatoire de
Haute Provence and to A. Baranne for their help in refurbishing the MARLY
telescope and remounting it in La Silla. We are also grateful for the support
given to our project by the technical staff at ESO, La Silla. We thank
J.F. Lecointe for assistance with the online computing.
We also thank S.~Deutsua, R.~McMillan, H.~Newberg, P.~Nugent, 
C.~Pennypacker, S.~Perlmutter and M.~Strauss for providing spectra of 
SN 1997eb, 
SN 1997ec, SN 1997ed and SN 1997ee. We are grateful to 
R.~Pain for helpful discussions. 
We also thank  the referee for his helpful remarks.    
\end{acknowledgements}


\begin{thebibliography}{}
   \bibitem[1984]{BAH84} Bahcall  J. N., Soneira R., 1984, A\&AS, 55, 67.  
    \bibitem[1997]{BAU97} Bauer F. et al. (EROS collaboration), 1997, 
in Proceedings of the ``Optical Detectors for Astronomy'' 
workshop (ESO, Garching)    
    \bibitem[1996]{BER96} Bertin E., Arnouts S., 1996, A\&AS, 117, 393.
    \bibitem[1997]{BER97} Bertin E., Dennefeld M., 1997, A\&A, 317, 43.
    \bibitem[1999]{CAP99} Cappellaro E., Evans R., Turatto M. et al., 
 1999, A\&A, 351, 459.
    \bibitem[1997]{CAP97} Cappellaro E., Turatto M., 
Tsvetkov, D. Yu  et al., 
 1997, A\&A, 322, 431.
     \bibitem[1999]{DER99} Derue F., Afonso C.,  Alard C.  et al., 1999,  A\&A, 351, 87.
   \bibitem[1997]{FIO97} Fioc M.,  Rocca-Volmerange B., 1997, A\&A, 326, 950.
  \bibitem[1996]{HAM96} Hamuy M., Phillips M.M., Suntzeff N.B.  et al., AJ, 112, 2391.
  \bibitem[1998]{HAR98} Hardin D., 1998, PhD thesis in French, Universit\'e de Paris 11, DAPNIA-SPP 98-1002.
  \bibitem[1997]{IAU6760} IAU circular 6760, 1997, the EROS collaboration.
    \bibitem[1997]{IAU6762} IAU circular 6762, 1997, the EROS collaboration.
    \bibitem[1997]{IAU6782} IAU circular 6782, 1997, the EROS collaboration.
    \bibitem[1997]{IAU6785} IAU circular 6785, 1997, Deustua S., McMillan R., Newberg H.J.M.  et al.    
    \bibitem[1998]{IAU7046} IAU circular 7046, 1998, Aldering G., the Supernova Cosmology Project.
   \bibitem[1992]{LAN92} Landolt A.U.,  1992, AJ, 104, 340.
   \bibitem[1996]{LIN96} Lin H.,  Kirshner R.P., Shectman S.A.  et al.,  1996, ApJ, 464, 60.
   \bibitem[1992]{MUL92} Muller R.A., Newberg H.J.M., Pennypacker C.R. 
 et al.,  1992, ApJ, 384, L9.
  \bibitem[1995]{NUG95} Nugent P.,  Phillips M.M., Baron E.  et al., 1995, ApJ, 455, L147.
  \bibitem[in preparation]{NUG99} Nugent P.,  Kim A., in preparation.
   \bibitem[1996]{PAI96} Pain R., Hook I.M., Deustua S.  et al., 1996, 473, 356.
  \bibitem[1998]{PAL98} Palanque-Delabrouille N., Afonso C., Albert J.-N.  et al., 1998, A\&A, 332, 1.
   \bibitem[1998]{PER97} Perlmutter S. et al., 1997, in 
  Thermonuclear supernov{\ae},
eds. R. Canal, P. Ruiz-Lapuente, J. Isern,  
Kluwer Academic Publishers, Dordrecht p. 749.
   \bibitem[1999]{PER99} Perlmutter S., Aldering G., Goldhaber G. et al., 1999, ApJ, 517, 565.
  \bibitem[1993]{PHI93} Phillips M.M., 1993, ApJ, 413, L105.
   \bibitem[1991]{PIC91} Picard A., 1991, AJ 102,  445.
  \bibitem[1998]{REI98} Reiss D.J., Germany L.M., Schmidt B.P. et al., 1998, AJ, 115, 26.
  \bibitem[1996]{RIE96} Riess A.G., Press W.H.,  Kirshner R. P., 1996, ApJ, 473, 88.
  \bibitem[1998]{RIE98} Riess A.G., Filippenko A.V., Challis P., 1998, AJ, 116, 1009.
   \bibitem[1998]{SCH98} Schlegel D. J., Finkbeiner D. P., Davis M., 1998, ApJ, 500, 525. 
    \bibitem[1996]{SHE96} Shectman S.A., Landy, S.D.,
 Oemler, A. et al., 1996, ApJ 470, 172.
    \bibitem[1995]{WEI95} Weir N., Djorgovski S., Fayyad U.M., 1995, AJ 110, 1. 



\end{thebibliography}
\end{document}